\documentclass[conference,keeplastbox]{IEEEtran} 
\usepackage{cite}
\usepackage{amsmath,amssymb,amsfonts}
\usepackage{algorithmic}
\usepackage{graphicx}
\usepackage{textcomp}
\usepackage{xcolor}
\usepackage{float}
\usepackage{subfigure}

\usepackage{hyperref}

\usepackage{graphicx}  

\usepackage{caption}
 
\DeclareCaptionLabelFormat{lc}{{#1}~#2}
\makeatletter
\def\endthebibliography{%
  \def\@noitemerr{\@latex@warning{Empty `thebibliography' environment}}%
  \endlist
}
\makeatother

 \usepackage{epsfig,endnotes}

\usepackage{floatpag,enumitem}

\usepackage{relsize}
 
 \usepackage{tikz}

\usepackage{multirow}
\usepackage{setspace}
\usepackage{mathrsfs}
\usepackage{bbm}
\usepackage{pifont}
\usepackage{amsfonts}

\usepackage{amsmath, amsthm}

\usepackage{tabularx}
\usepackage{listings}
\usepackage{graphicx}
\usepackage{amssymb,booktabs}
\usepackage{array}
\usepackage{cases}

 \usepackage{supertabular}
\usepackage{mathrsfs}

\usepackage{psfrag}
\usepackage{bbding}

\usepackage{float}

\usepackage{amsbsy}

\makeatletter
\providecommand{\leftsquigarrow}{%
  \mathrel{\mathpalette\reflect@squig\relax}%
}
\newcommand{\reflect@squig}[2]{%
  \reflectbox{$\m@th#1\rightsquigarrow$}%
}
\makeatother

\usepackage{algorithm}
 \usepackage{algorithmic}

\makeatletter
\newcommand{\newalgname}[1]{%
  \renewcommand{\ALG@name}{#1}%
}

\newcommand {\C} {{\rm I\kern-5.5pt C}}

\def\centerhack#1{\hbox to 0pt{\hss\footnotesize #1\hss}}
\def\centerhackn#1{\hbox to 0pt{\hss #1\hss}}
\def\dchack#1{\vbox to 0pt{\vss{\hbox to 0pt{\hss#1\hss}}\vss}}

\usepackage{etoolbox}
\AtBeginEnvironment{align}{\setcounter{subeqn}{0}}
\newcounter{subeqn} %

\usetikzlibrary{positioning}

\newcounter{mysub}

\newtheorem*{proposition1.1}{Proposition 1.1}
\newtheorem*{proposition1.2}{Proposition 1.2}
\newtheorem*{proposition1.3}{Proposition 1.3}
\newtheorem*{proposition2.1}{Proposition 2.1}
\newtheorem*{proposition2.2}{Proposition 2.2}
\usepackage{subfigure}

\begin{document}


\title{A Survey of Intelligent Reflecting Surfaces (IRSs): Towards 6G Wireless Communication Networks}


\author{\IEEEauthorblockN{Jun Zhao}
\IEEEauthorblockA{Assistant Professor\\
Nanyang Technological University, Singapore
\\
JunZhao@ntu.edu.sg
} \and
\IEEEauthorblockN{Yang Liu}
\IEEEauthorblockA{Research Fellow\\Nanyang Technological University, Singapore
\\
liuocean613@gmail.com
}}

\markboth{Journal of \LaTeX\ Class Files,~Vol.~14, No.~8, August~2015}%
{Shell \MakeLowercase{\textit{et al.}}: Bare Demo of IEEEtran.cls for IEEE Journals}
%



\maketitle

 \pagestyle{plain} \thispagestyle{plain}

\begin{abstract}
Intelligent reflecting surfaces (IRSs) tune wireless environments to increase spectrum and energy efficiencies. In view of much recent attention to the IRS concept as a promising technology for 6G wireless communications, we present a survey of IRSs in this paper. Specifically, we categorize recent research studies of IRSs as follows. For IRS-aided communications, the summary includes capacity/data rate analyses, power/spectral optimizations, channel estimation, deep learning-based design, and reliability analysis. Then we review IRSs implementations as well as the use of IRSs in secure communications, terminal-positioning, and other novel applications. We further identify future research directions for IRSs, with an envision of the IRS technology playing a critical role in 6G communication networks similar to that of massive MIMO in 5G networks. As a timely summary of IRSs, our work will be of interest to both researchers and practitioners working on IRSs for 6G networks. \end{abstract}


\section{Introduction}

As 5G communication networks are now putting into commercialization~\cite{patzold2018s}, technologies for the next-generation (i.e., 6G) communications are also being explored to achieve faster and more reliable data transmissions~\cite{saad2019vision}. Among these technologies, intelligent reflecting surfaces have received much interest recently in the academia~\cite{huang2019Reconfigurable,hu2018beyond,hu2018beyondpositioning,cui2019secure} and industry~\cite{DOCOMO}. In November 2018, the Japanese mobile carrier NTT DoCoMo and a smart radar startup MetaWave demonstrated the application of meta-structure
technology to data communication in 28GHz band~\cite{DOCOMO}.

\textbf{Intelligent reflecting surfaces.} An intelligent reflecting surface (IRS)~\cite{huang2019Reconfigurable} comprises an array of IRS units, each of which can independently incur some change to the incident signal~\cite{basar2019wireless}. The change in general may be about the phase, amplitude, frequency, or even polarization~\cite{basar2019wireless}. To date, in most studies~\cite{jung2018performance,hu2018beyond,hu2017potential,guo2019weighted,huang2019Reconfigurable,yu2019miso,jung2019performance}, the change is considered as a phase shift only to the incident signal, so that an IRS consumes \textit{no} transmit power. In essence, an IRS intelligently configures the wireless environment to help the transmissions between the sender and receiver, when direct communications have bad qualities. Example places to put IRSs are walls, building facades, and ceilings~\cite{qingqing2019towards}.

\begin{figure}[!t]
 \centering
\includegraphics[scale=0.3]{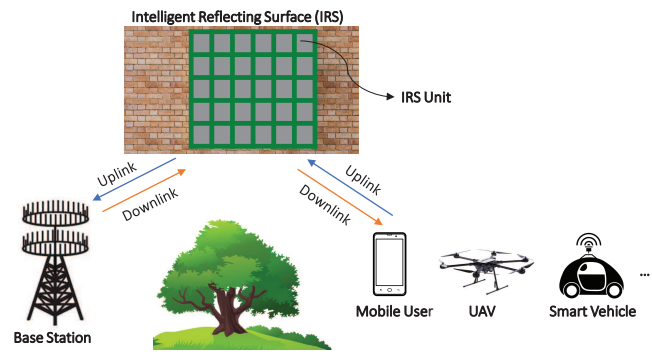} 
\caption{Communications between a base station and a mobile user/UAV/smart vehicle with the aid of an intelligent reflecting surface (IRS).} \label{fig-IRS}
\end{figure}

\textbf{IRS-aided communications.} Figure~\ref{fig-IRS} illustrates IRS-aided communications between a base station (BS) and a mobile user (MU), an unmanned aerial vehicle (UAV), a smart vehicle, or any other terminal, where a tree blocks the line-of-sight. In the rest of the paper, we mostly consider IRS-aided communications of a BS and a MU or several MUs without loss of generality.

\textbf{Comparing IRSs with massive MIMO and other related technologies.} The IRS concept  can be considered to be related with the technology of massive \mbox{multiple-input multiple-output (MIMO)}~\cite{hu2018beyond}, where large arrays of antennas are utilized to improve spectral and energy efficiencies. Hence, we envision IRS to play a crucial role in 6G communication networks similar to that of massive MIMO in 5G networks. Thus, IRS can be used to help achieve massive MIMO 2.0~\cite{sanguinetti2019towards}.
What differentiates IRSs from massive MIMO is that IRSs tune the wireless propagation environment for the communication~\cite{guo2019weighted,Liaskos2018}. Hu~\textit{et~al.}~\cite{hu2017potential} present the first analysis on information-transfer capabilities of IRSs. They prove that the capacity which can be harvested per square metre ($m^2$) surface-area has a linear relationship with the average transmit power, instead of being logarithmic in the case of a massive-MIMO deployment. In addition to massive MIMO, other existing technologies which have been compared with IRSs in recent studies~\cite{wu2018intelligent,di2019smart2} include backscatter communication~\cite{parks2015turbocharging}, millimeter (mm)-wave communication~\cite{rappaport2015wideband},
and network densification~\cite{andrews2016we}. However, these related technologies do not control the wireless environment and typically consume much power~\cite{nadeem2019intelligent}.

\textbf{Comparing IRS with other recent notions.} In addition to IRS, other recent notions have also been given in the literature. These names include the following:
\begin{itemize}
\item \textit{large intelligent surface (LIS)}
~\cite{hu2018beyond, hu2018beyondpositioning, hu2017potential, hu2018capacity, hu2017cramer,jung2018performance, jung2019performance, Junginproceedings, jung2019reliability,nadeem2019large}, which is preferred for asymptotic analysis assuming infinite length(s) of the surface, or sufficiently large number of antennas. It is worth noting that in \cite{hu2018beyond, hu2018beyondpositioning, hu2017potential, hu2018capacity, hu2017cramer,jung2018performance, jung2019performance, Junginproceedings, jung2019reliability}, the surface indeed transmits signals actively, instead of passively reflecting signals from base stations as in the case of IRS.  

\item \textit{large intelligent metasurface (LIM)}~\cite{he2019cascaded} and \textit{programmable/reconfigurable metasurface}~\cite{di2019reflection,MetaSurface2018,MetaSurface2019,MetaSurfaceLiquid}, where the prefix ``meta-'' derives from a metallic pattern called \emph{meta-atom}, through which the surface is engineered to have some property which is not found in naturally occurring surfaces
\item 
 \textit{smart reflect-arrays}~\cite{tan2018enabling,tan2016increasing,nie2019intelligent}, which emphasize the surface's reflection function (in the same spirit as IRS), instead of being used for transmission that is provided by amplify-and-forward/decode-and-forward relays such as MIMO technologies,
\item \textit{reconfigurable intelligent surface}~\cite{huang2018Reconfigurable,huang2018largeenergy,basar2019wireless}, where ``reconfigurable'' means that the
angle of reflection can be reconfigured (via software) regardless of the angle of incidence,
\item \textit{software-defined surface
(SDS)}~\cite{basar2019large} and \textit{software-defined metasurfaces (SDMs)}~\cite{liaskos2019interpretable},
 which are inspired by
the definition of software-defined radio~\cite{ulversoy2010software} and consider the interaction between the surface and incoming waves to have a software-defined fashion,
\item \textit{passive intelligent surface (PIS)}~\cite{mishra2019channel} and \textit{passive intelligent mirrors}~\cite{huang2018achievable},
which underline the passive reflection without consuming transmit power.
\end{itemize} 
For consistency, in the rest of the paper, we will use the name IRS instead of other terms listed above. In addition, it is worth mentioning that frequency selective surfaces recently studied in~\cite{zhang2018graphene,zhang2019mutual} are used to reduce the coupling effects in ultra-massive
MIMO and are different from IRSs.

\textbf{Organization of this paper.} In Section~\ref{sec-Categorizing}, we classify recent studies of IRSs into different categories. Section~\ref{sec-other-survey} highlights the differences between our work and recent reviews
 of IRSs. In Section~\ref{sec-future}, we identify several directions for future research of IRSs. 
  Finally, Section~\ref{sec-Conclusion} concludes the paper.

\section{Categorizing Recent Studies of Intelligent Reflecting Surfaces} \label{sec-Categorizing}

We now classify recent research work on IRSs. For IRS-aided communications, we discuss capacity/data rate analyses, power/spectral optimizations, channel estimation, deep learning-based design, and reliability analysis. We also  review IRSs implementations as well as the use of IRSs in secure communications, terminal-positioning, and other novel applications.
 
\subsection{Capacity/data rate analyses of IRS-aided communications} \label{subsection-Capacity}

Hu~\textit{et~al.}~\cite{hu2017potential} establish that the capacity achieved per square metre ($m^2$) surface-area is linearly proportional with the average transmit power, instead of having a logarithmic relationship as the case of massive MIMO.

Hu\textit{~et~al.}~\cite{hu2018beyond} analyze capacities of single-antenna terminals communicating to an IRS. They first consider the entire surface as a receiving antenna array. In this setting, for a sufficiently large surface-area, their result is that the received signal following a matched-filtering operation can be well represented by a sinc-function-like intersymbol interference channel. Afterwards, they derive the capacity per square metre ($m^2$) surface-area and show its convergence to $\frac{P}{2 N_0}$~[nats/s/Hz/volume-unit] when the wavelength $\lambda$ tends to zero, where $P$ is the transmit power per volume-unit (which can be $m$, $m^2$, or $m^3$), and $N_0$ denotes the additive white Gaussian noise's 
 power spectral density.

A recent work~\cite{hu2018capacity} by Hu~\textit{et~al.} examines the degradations in capacity when IRSs are allowed to have hardware impairments. They show that splitting an IRS into an array consisting of a number of small IRS units can mitigate the degradation.

The conference paper~\cite{Junginproceedings} and its full version~\cite{jung2018performance}
by Jung\textit{~et~al.} present an asymptotic analysis of the uplink data rate in an IRS-based communication system. 
 Their analysis considers channel estimation errors and model interference channels to be spatially correlated Rician fading~\cite{Hamdi2008CapacityOM}. Furthermore, channel hardening effects are also taken into consideration. They show that the asymptotic capacity result is in accordance with  the exact mutual information as the numbers of antennas and mobile devices increase. 
For uplink data rates in IRSs, Jung~\textit{et~al.}~\cite{Junginproceedings} present an asymptotic analysis where channel estimation
errors are taken into consideration.

 Guo~\textit{et~al.}~\cite{guo2019weighted} maximize the weighted sum of downlink rates by finding the optimal active beamforming at the BS and  passive
beamforming at the IRS, where each weight represents the priority of a mobile user. For practicality and simplicity of optimization analysis, they consider the IRS phase shifts to take only discrete values.

Nadeem~\textit{et~al.}~\cite{nadeem2019large} consider a single-cell multi-user system, where a base station (BS) with multiple antennas communicates many single-antenna users via an IRS. For the downlink, they investigate how to maximizes the SINR by optimizing the linear precoder and power
allocation at the BS, and the IRS phase matrix. Their analysis involves different rank structures of the channel matrix between the BS and
the IRS, and also the spatial correlations among the IRS elements.

The conference paper~\cite{wu2018intelligent} and its full version~\cite{wu2018intelligentfull} by Wu~\textit{et~al.} maximize the signal-to-interference-plus-noise ratio (SINR) received at mobile users by jointly optimizing the IRS phase matrix and the transmit beamforming of the active antenna array at the BS.

\subsection{Power/spectral optimizations in IRS-aided communications}
 
In this subsection, we summarize recent optimization studies of power/spectral efficiency in IRS-aided communications.
 
The work~\cite{huang2019Reconfigurable} and its earlier version~\cite{huang2018largeenergy} by Huang~\textit{et~al.} maximize the \textit{bit-per-Joule energy efficiency} of the downlinks by finding the IRS phase matrix and the optimal power
allocation at the BS. To simplify the analysis, they consider that the BS employs a well-designed zero-forcing precoding matrix to achieve perfect interference suppression among signals received by the mobile users.

Fu~\textit{et~al.}~\cite{fu2019intelligent} solve the downlink transmit power minimization for an IRS-aided multiple
access network by optimizing both the transmit
beamformers at the BS and the phase shift matrix at the IRS.

Yu~\textit{et~al.}~\cite{yu2019miso} and Jung~\textit{et~al.}~\cite{jung2019performance} both consider maximization problems of the
\textit{spectral efficiency} in IRS-aided communication systems. Specifically, the work~\cite{yu2019miso} maximizes the spectral efficiency by optimizing the beamformer at the access point and the IRS phase shifts. The study~\cite{jung2019performance} considers the typical setting where pilot signaling is used to obtain channel state information and hence pilot training structure impacts the achievable spectral efficiency. The authors of~\cite{jung2019performance} first derive an asymptotic value of the spectral efficiency and then use the result to find the optimal pilot training length which maximizes the asymptotic spectral efficiency.

\subsection{Channel estimation for IRS-aided communications} \label{subsection-Channel-estimation}

In a typical setting, 
IRSs are passive and do not have sensing capabilities, so downlinks are estimated at the base station via control signals. Then the channel information is reported by the base station to the IRS controller, which sets the phase shifts accordingly~\cite{nadeem2019intelligent}.  

Nadeem~\textit{et~al.}~\cite{nadeem2019intelligent} present a channel estimation protocol based on minimum mean squared error (MMSE). Specifically, they divide the total channel estimation time into a number of sub-phases. In the first sub-phase, all IRS units are turned OFF and the base station estimates the direct channels for all users.  In each of the following sub-phases, each IRS element takes turns to be ON while all other IRS units are OFF, to allow estimations by the base station. At the end of the protocol, estimation results of all sub-phases are taken together using the MMSE approach to obtain a comprehensive picture of channel estimation.

Taha~\textit{et~al.}~\cite{taha2019enabling} address the channel estimation problem using \textit{compressive sensing}~\cite{Gao2018CompressiveST} and \textit{deep learning}~\cite{Zhang2019Deep}. In their setting, IRS units which are connected to the baseband of the IRS controller are referred to as being \textit{active}, and the rest IRS units are considered as \textit{passive}. In the proposed solutions, they utilize compressive sensing and deep learning techniques respectively to estimate the channels at all the IRS units from the channels seen only at the active IRS units. 
 
He~\textit{et~al.}~\cite{he2019cascaded} tackle channel estimation for IRS-aided MIMO systems using a three-stage mechanism. The three stages include sparse matrix factorization, ambiguity elimination, and matrix completion, respectively. The first stage takes the received signal and uses matrix factorization to derive the channel
matrix between the base station and the IRS, as well as the channel
matrix between the IRS and the mobile user. The second stage eliminates the ambiguity of the solutions to matrix factorization, using the information of the IRS state matrix, which contains the ON/OFF information of each IRS unit at each time. The third stage uses properties of the channel
matrices to recover the missing entries. The three stages are solved by the algorithms of bilinear generalized approximate message passing~\cite{Parker2014BilinearGA}, greedy pursuit~\cite{Cheng2012Compressed}, and Riemannian manifold gradient~\cite{Knuth2013CollaborativeLW}, respectively. 

Mishra and Johansson~\cite{mishra2019channel} design a channel estimation mechanism for IRS-aided
energy transfer from a power beacon with multiple antennas to
a single-antenna user. They further use the estimation results to design active and passive energy
beamforming at the power beacon and IRS, respectively. Independent of~\cite{nadeem2019intelligent}, the work~\cite{mishra2019channel} also proposes the approach of dividing the total channel estimation time into several sub-phases and allowing only one IRS unit to be ON in each sub-phase. Moreover, the channel estimation protocol in~\cite{mishra2019channel} also permits that ON/OFF modes of IRS units may not be implemented perfectly in practice.

As already discussed in Section~\ref{subsection-Capacity}, Jung\textit{~et~al.}~\cite{Junginproceedings,jung2018performance} take channel estimation errors into account to analyze uplink data rates of IRS-aided communications.

\subsection{Deep learning-based design for IRS-aided communications}

Liaskos\textit{~et~al.}~\cite{liaskos2019interpretable} use deep learning for configuring IRSs to aid wireless communications. Specifically, they regard wireless propagation as a deep neural network, where IRS units are neurons and their cross-interactions
as links. After training from the data, the wireless network learns the propagation basics of IRSs and configures them to the optimal setting.

As already discussed in Section~\ref{subsection-Channel-estimation}, 
Taha~\textit{et~al.}~\cite{taha2019enabling} utilize deep learning for channel estimation in IRS-aided communications. Specifically, qualities of wireless channels at all the IRS units are learnt via a deep neural network using channels seen only at those IRS units which are connected to the baseband of the IRS controller. Furthermore, deep learning is used to guide the IRS to learn the optimal interaction with the incident signals. A short conference version of~\cite{taha2019enabling} appears as~\cite{taha2019deep}. 
 
\subsection{Reliability analysis of IRS-aided communications}

Jung\textit{~et~al.}~\cite{jung2019reliability} present a reliability analysis of IRS-aided communications in terms of uplink rate distribution and outage probability. The distribution of the data sum-rate is obtained using the Lyapunov central limit theorem~\cite{chafii2016necessary}. Then the outage probability is given by the probability that the sum-rate is below a desired value. Note that although the authors' earlier work~\cite{jung2018performance} analyzes the mean and variance of the rate, the probabilistic distribution of the rate is needed to compute the outage probability.

\subsection{Implementations of IRSs}

Hu\textit{~et~al.}~\cite{hu2018beyond} investigate IRS implementations as a grid of antenna elements.
 Subject to the constraint that every spent antenna can earn one signal space dimension, they show that the hexagonal lattice minimizes the IRS surface-area given
  the desired number of independent 
 signal dimensions. 
 The analysis of~\cite{hu2018beyond} leverages the classical lattice theory~\cite{ganter1997applied}, which has various applications beyond wireless communication, including information theory~\cite{damen2003maximum}, cryptography~\cite{sotiraki2018ppp}, machine learning~\cite{liquiere1998structural}, and knowledge representation~\cite{wu2008granular}.
 
 Taha~\textit{et~al.}~\cite{taha2019enabling,taha2019deep} propose IRS architectures consisting of two types of IRS units: \textit{active} and \textit{passive} ones. An active IRS unit is connected to the baseband of the IRS controller, whereas a passive one is not. Then the system optimization can be booted from channel information at the active IRS units, which capture the environmental conditions and sender/receiver locations.

\subsection{IRSs for secure communications}

A number of recent studies~\cite{yu2019enabling,chen2019intelligent,shen2019secrecy,cui2019secure} have leveraged IRSs to secure the physical layer of wireless communications. In the simplest wiretap channel introduced first by Wyner~\cite{wyner1975wire}, a transmitter and a
legitimate receiver have communications, which are wiretapped by an eavesdropper. This simple model has been extended to the broadcast wiretap channel~\cite{csiszar1978broadcast}, compound wiretap channel~\cite{liang2009compound},
Gaussian wiretap channel~\cite{leung1978gaussian}, and 
MIMO wiretap channel~\cite{oggier2011secrecy}.

The intuition of applying an IRS to secure communications under a wiretap channel is that an IRS can be used to increase the data rate at a legitimate receiver while decreasing the data rate at an eavesdropper. This improves the difference between the two rates (the former minus the latter), which is defined as the secrecy data rate. We now elaborate the contributions of~\cite{yu2019enabling,chen2019intelligent,shen2019secrecy,cui2019secure}.

Cui~\textit{et~al.}~\cite{cui2019secure}, Shen~\textit{et~al.}~\cite{shen2019secrecy}, and Yu~\textit{et~al.}~\cite{yu2019enabling} study an IRS-aided 
 communication system with a multi-antenna
transmitter communicating to a single-antenna legitimate receiver in the presence of an eavesdropper. All of the three papers~\cite{cui2019secure,yu2019enabling,shen2019secrecy} consider the optimal design of the base station's transmit beamforming and the IRS's reflect beamforming to maximize the legitimate communication link's secrecy rate subject to the base station's transmit
power constraint. The differences among~\cite{cui2019secure,yu2019enabling,shen2019secrecy} lie in the specific details of the approaches to solving the optimization problems. In particular,~\cite{cui2019secure} uses \textit{alternating optimization}~\cite{liu2014multi} to design the transmit beamforming and IRS phase shifts alternately. More specifically, in each iteration of~\cite{cui2019secure}, the transmit beamforming can be exactly computed, but for deriving the IRS phase shifts, the \textit{semidefinite relaxation}~\cite{luo2010semidefinite} technique and the \textit{Charnes--Cooper transformation}~\cite{naeem2013energy} are combined to convert a non-convex problem into a convex \textit{semidefinite programming} problem~\cite{ouyang2010received}, which can be solved by the interior-point method~\cite{xu2011source}. In each iteration of~\cite{yu2019enabling}, the transmit beamforming is determined by a generalized eigenvalue problem~\cite{nadakuditi2010fundamental}, while each phase shift is solved by an element-wise \textit{block coordinate descent} method~\cite{wu2018joint}, which can be seen as a generalization of \textit{alternating optimization} and optimizes the objective function with respect to a block of
optimization variables in each iteration while fixing the other blocks. Finally, the work~\cite{shen2019secrecy} also adopts \textit{alternating optimization} to compute transmit beamforming and the IRS phase shifts alternately, where in each iteration the former can be exactly derived and the latter is solved by letting the objective function be its approximation using results from~\cite{sun2016majorization}.

Different from~\cite{cui2019secure,shen2019secrecy,yu2019enabling} above, Chen~\textit{et~al.}~\cite{chen2019intelligent} examine the case of multiple legitimate receivers and multiple eavesdroppers. Specifically, in~\cite{chen2019intelligent}, the considered IRS-aided downlink
 broadcast system consists of a multi-antenna
base station, multiple legitimate receivers with each having single antenna, and multiple eavesdroppers. Then~\cite{chen2019intelligent} maximizes the minimum secrecy data
rate among all legitimate receivers by finding the optimal transmit beamforming and IRS phase shifts via \textit{alternating optimization}, where both cases of phase shifts taking discrete and continuous values are considered. Moreover,~\cite{chen2019intelligent} also studies a case where the IRS reflecting amplitude is allowed to be less than $1$. The optimization techniques used by~\cite{chen2019intelligent} include \textit{path-following iterative} algorithms~\cite{nasir2016path} and heuristic projection methods~\cite{ferreira2015projection}.

\subsection{IRSs for terminal-positioning and other novel applications}

The conference paper~\cite{hu2017cramer} and its journal version~\cite{hu2018beyondpositioning} by Hu~\textit{et~al.} utilize IRS for terminal-positioning. In particular, they derive the Cram{\'e}r--{Rao} lower bounds (CRLBs)~\cite{angjelichinoski2015cramer} for all three Cartesian dimensions of a terminal. The result is that in general  the CRLB decreases quadratically with respect to the IRS surface-area, except for the case of a terminal locating perpendicular to the IRS center where the CRLB for the distance from the IRS decreases   linearly in the IRS surface-area. The analyses in~\cite{hu2017cramer,hu2018beyondpositioning} also involve the Fisher information, since it is no greater than the CRLB of an unbiased estimator~\cite{zhao2016er}.

Basar~\cite{basar2019large} uses IRS to aid index modulation (IM), which manipulates the indices of the transmit entities to convey information~\cite{basar2017index}. In~\cite{basar2017index}, with IRS-space shift keying and IRS-spatial modulation, IM is realized by selecting a particular receive antenna index based on the information bits.

Jiang~\textit{et~al.}~\cite{jiang2019over} employ IRS to assist \textit{over-the-air computation} (AirComp), where the base station aims to compute some function from data of all mobile users. The optimization  problem formulated in~\cite{jiang2019over} is find the IRS phase shifts and the base station' decoding vectors to minimize the distortion after signal decoding, which is defined as the mean squared error among the decoding results. In view of the non-convexity of the problem, the authors of~\cite{jiang2019over} use the majorization-minimization~\cite{sun2016majorization}
technique to propose an alternating difference-of-convex algorithm~\cite{zhang2018minimization}.

Basar~\cite{basar2019transmission}
proposes the novel concept
of using an IRS as an access point (AP). In the proposed design, an radio frequency (RF) signal generator close to the IRS generates an unmodulated carrier signal and sends it to the IRS with negligible fading. Then the adjustable IRS phase shifts are used to convey information bits. 

Mishra and Johansson~\cite{mishra2019channel} leverage IRS to support wireless energy transfer from a power beacon with multiple antennas to
a single-antenna user. As discussed previously in Section~\ref{subsection-Channel-estimation},~\cite{mishra2019channel} first presents a channel estimation protocol and then uses the estimation results to set energy beamforming at the power beacon and IRS, respectively. 

 \section{Recent Reviews of Intelligent Reflecting Surfaces and Related Technologies} \label{sec-other-survey}

Three recent studies~\cite{basar2019wireless,di2019smart2,qingqing2019towards} provide nice overviews for the IRS technology. Compared with our current paper, Basar~\textit{et~al.}~\cite{basar2019wireless} elaborate many mathematical details in IRS-aided communications. Di Renzo~\textit{et~al.}~\cite{di2019smart2} highlight the flavor of artificial intelligence in IRSs for empowering \textit{smart radio environments}. Wu~\textit{et~al.}~\cite{qingqing2019towards} focus on the key challenges in the design and implementation of IRS-aided communications.  
 Compared with our current paper,~\cite{basar2019wireless,di2019smart2,qingqing2019towards} do not thoroughly categorize IRS studies appearing in the literature. 
 Moreover, recent papers which are not covered by~\cite{basar2019wireless,di2019smart2,qingqing2019towards} but are discussed in our current work include \cite{cui2019secure,fu2019intelligent,guo2019weighted,hu2017cramer,hu2018capacity,Junginproceedings,nadeem2019intelligent,shen2019secrecy,yu2019miso}.

Other recent reviews are listed as follows. Tasolamprou~\textit{et~al.}~\cite{Tasolamprou2019arXivExploration} present a detailed discussion about issues in physical implementations of IRSs for wireless millimeter (mm)-wave communications. Sanguinetti~\textit{et~al.}~\cite{sanguinetti2019towards} summarize techniques for improving massive MIMO to its 2.0 version, where the IRS technology is only briefly mentioned. 
   Bj{\"o}rnson~\textit{et~al.}~\cite{bjornson2019massive} discuss different potential technologies related to massive MIMO, including intelligent massive MIMO, large-scale MIMO radar, and holographic massive MIMO.

\section{Future Research Directions for Intelligent Reflecting Surfaces} \label{sec-future}
 
We envision that IRS will play a fundamental role in 6G wireless communication networks similar to that of massive MIMO in 5G networks. Moreover, IRS can be utilized to help realize massive MIMO 2.0~\cite{sanguinetti2019towards}. Now
we identify three  future  directions for IRS research.

First, most existing studies on IRSs and their applications to wireless communications are about theoretical analyses with simulations as validations. Hence, an important research direction is to confirm the theoretical results with data from real-world system implementations and experiments. 

Second, existing models on how IRSs change the incident signals are simple. In contrast, an IRS's behavior depends on its physical materials and manufacturing processes~\cite{qingqing2019towards}. Models taking these issues into account can more accurately guide the optimization of IRSs for aiding wireless communications.

Third,     scaling laws need to be established for a fundamental understanding of the performance limits in IRS-aided communications. Answering this question requires a deep understanding of how IRSs impact traditional information-theoretic
models.
\section{Conclusion} \label{sec-Conclusion}

In this survey paper, we categorize recent studies of IRSs and identify future research directions for IRSs.
 As a promising technology to facilitate 6G wireless communications, IRSs induce 
smart radio environments to increase spectrum and energy efficiencies. We envision that the coming   years will see much research and development for the IRS technology to build 6G communications.



 
 

\end{document}